\documentclass{article}
\usepackage{amsfonts}
\usepackage{amsmath}
\usepackage{graphicx}
\usepackage{cite}
\usepackage{hyperref}
\usepackage{booktabs}

\begin{document}

\title{Asymmetric Duffing oscillator: the birth and build-up of period doubling}
\author{Jan Kyzio{\l},  Andrzej Okni\'{n}ski \\
Politechnika \'{S}wi\c{e}tokrzyska, Al. 1000-lecia PP 7,\\
 25-314 Kielce, Poland}
\maketitle

\begin{abstract}
In this work, we investigate the period doubling phenomenon in the
periodically forced asymmetric Duffing oscillator. We use the known steady-state
asymptotic solution -- the amplitude-frequency implicit function -- and
known criterion for the existence of period doubling. Working in the framework of
differential properties of implicit functions we derive 
analytical formulas for the  birth of period-doubled solutions. 
\end{abstract}

\section{Introduction and motivation}

\label{introduction} A period-doubling cascade of bifurcations is a typical
route to chaos in nonlinear dynamical systems. We shall study this
phenomenon in Duffing-type oscillator equations.

In this work, we study period doubling in the forced asymmetric Duffing
oscillator governed by the non-dimensional equation:%
\begin{equation}
\ddot{y}+2\zeta \dot{y}+\gamma y^{3}=F_{0}+F\cos \left( \Omega t\right) ,
\label{AsymDuffing}
\end{equation}%
which has a single equilibrium position and a corresponding one-well
potential \cite{Kovacic2011}, where $\zeta $, $\gamma $, $F_{0}$, $F$ are
parameters and $\Omega $ is the angular frequency of the periodic force.

The period doubling scenario in the dynamical system (\ref{AsymDuffing}) was investigated by 
Szempli\'{n}ska-Stupnicka in a series of groundbreaking papers \cite
{Szemplinska1986,Szemplinska1987,Szemplinska1988}, see also \cite
{Kovacic2011} for a review and further results.

The main idea introduced in \cite{Szemplinska1986} consists of perturbing the
main steady-state asymptotic solution of Eq.(\ref{AsymDuffing}), a $1:1$ resonance: 
\begin{equation}
y_{0}\left( t\right) =A_{0}+A_{1}\cos \left( \Omega t+\theta \right) ,
\label{AD-solution}
\end{equation}
as 
\begin{subequations}
\label{PERTURBED}
\begin{eqnarray}
y\left( t\right)  &=&y_{0}\left( t\right) +u\left( t\right) ,  \label{y} \\
u\left( t\right)  &=&B\cos \left( \tfrac{1}{2}\Omega t+\varphi \right) ,
\label{u}
\end{eqnarray}
\end{subequations}
substituting $y\left( t\right) $ into Eq.(\ref{AsymDuffing}) and considering
the condition of non-zero solution (\ref{u}). In papers \cite%
{Kovacic2011,Szemplinska1986,Szemplinska1987,Szemplinska1988} several
conditions guaranteeing the formation and stability of solution (\ref{PERTURBED}) were
found and used to study the period doubling phenomenon.

For example, these authors were able to find intervals $\left( \Omega _{1},\
\Omega _{2}\right) $ in which solution (\ref{AD-solution}) destabilized with
the formation of period-doubled solution (\ref{u}) \cite{Kovacic2011}, and
demonstrate that a cascade of period doubling leading to chaos was formed 
\cite{Kovacic2011,Szemplinska1986,Szemplinska1987,Szemplinska1988}.

Our motivation is fueled by the observation that steady-state solution (\ref%
{AD-solution} as well as period-doubling conditions found in \cite%
{Kovacic2011,Szemplinska1986,Szemplinska1987,Szemplinska1988} are in the form of
some implicit functions. Therefore, it should be possible, within the
framework of differential properties of implicit functions \cite{Kyziol2022},
to obtain new results concerning the period-doubling mechanism.

The aim of the present work is thus to apply this formalism to implicit
functions derived in \cite%
{Kovacic2011,Szemplinska1986,Szemplinska1987,Szemplinska1988}.

The paper is organized as follows. In Section \ref{steady} steady-state
solution of Eq.(\ref{AsymDuffing}) of form (\ref{AD-solution}) is reviewed
and a period doubling condition derived in \cite%
{Kovacic2011,Szemplinska1986,Szemplinska1987,Szemplinska1988} is described
in Section \ref{conditions}. In Section \ref{differential} we derive new
results concerning period doubling applying the formalism of differential
properties of implicit functions and in Section \ref{verification} we verify our results. 
In Section \ref{summary} we summarize our findings.

\section{The main resonance: steady-state solution}
\label{steady}
The steady-state solution of Eq.(\ref{AsymDuffing}) of form (\ref%
{AD-solution}), describing $1:1$ resonance, was computed in Refs. \cite%
{Szemplinska1986,Jordan1999,Kovacic2011}. Proceeding as in \cite{Kyziol2023}
we get two implicit equations for $A_{0},\ A_{1}$, and $\Omega $: 
\begin{subequations}
\label{1d1e}
\begin{eqnarray}
A_{1}^{2}\left( 3\gamma A_{0}^{2}+\frac{3}{4}\gamma A_{1}^{2}-\Omega
^{2}\right) ^{2}+4\Omega ^{2}\zeta ^{2}A_{1}^{2} &=&F^{2},  \label{1d} \\
\gamma A_{0}^{3}+\frac{3}{2}\gamma A_{0}A_{1}^{2}-F_{0} &=&0.  \label{1e}
\end{eqnarray}
Computing $A_{1}^{2}$ from Eq.(\ref{1e}) for $A_{0}\neq 0$ and substituting
into (\ref{1d}), we obtain finally one implicit equation for $A_{0},\ \Omega 
$ \cite{Kovacic2011,Kyziol2023}: 
\end{subequations}
\begin{equation}
f\left( A_{0},\Omega ;\gamma ,\zeta ,F_{0},F\right)
=\sum\nolimits_{k=0}^{9}c_{k}A_{0}^{k}=0,  \label{f}
\end{equation}
where the coefficients $c_{k}$ are given in Table \ref{tab:T1} (cf.
Eq.(8.3.12) in \cite{Kovacic2011}). \vspace{-0.2cm} 
\begin{table}[th]
\caption{Coefficients $c_{k}$ of the polynomial (\protect\ref{f})}
\label{tab:T1}\centering
\begin{tabular}{|l|l|}
\hline
$c_{9}=25\gamma ^{3}$ & $c_{4}=16\Omega ^{2}\gamma F_{0}$ \\ \hline
$c_{8}=0$ & $c_{3}=-9\gamma F_{0}^{2}+6\gamma F^{2}$ \\ \hline
$c_{7}=-20\Omega ^{2}\gamma ^{2}$ & $c_{2}=-4F_{0}\Omega ^{4}-16\zeta
^{2}\Omega ^{2}F_{0}$ \\ \hline
$c_{6}=-15\gamma ^{2}F_{0}$ & $c_{1}=4\Omega ^{2}F_{0}^{2}$ \\ \hline
$c_{5}=4\gamma \Omega ^{2}\left( \Omega ^{2}+4\zeta ^{2}\right)$ & $%
c_{0}=-F_{0}^{3}$ \\ \hline
\end{tabular}%
\end{table}

We can also obtain an implicit equation for $A_{1},\ \Omega$ as in \cite%
{Kyziol2023}. Solving Eq. (\ref{1e}) for $A_{0}$ (there is only one real
root) and substituting to (\ref{1d}) we get: 
\begin{equation}
g\left( A_{1},\Omega ;\gamma ,\zeta ,F,F_{0}\right) =A_{1}^{2}\left( 3\gamma
A_{0}^{2}+\tfrac{3}{4}\gamma A_{1}^{2}-\Omega ^{2}\right) ^{2}+4\Omega
^{2}\zeta ^{2}A_{1}^{2}-F^{2}=0,  \label{A1-b}
\end{equation}
where $A_{0}$ and $Y$ are defined as: 
\begin{equation}
A_{0}=-\frac{A_{1}^{2}}{2Y}+Y,\quad Y=\sqrt[3]{\sqrt{\frac{1}{8}A_{1}^{6}+%
\frac{1}{4\gamma ^{2}}F_{0}^{2}}+\frac{1}{2\gamma }F_{0}}.  \label{A1-a}
\end{equation}

\section{Birth of period doubling}

\label{conditions} 
The stability of the steady-state solution $y_{0}\left(
t\right) =A_{0}+A_{1}\cos \left( \Omega t+\theta \right) $ is studied via a
substitution in Eq. (\ref{AsymDuffing}): $y=y_{0}+u\left( t\right) $, with $%
u\left( t\right) $ small. This substitution leads to the (linear) Hill's
equation for $u\left( t\right) $ \cite{Kovacic2011,Szemplinska1986}: 
\begin{subequations}
\label{HILL-1}
\begin{gather}
\hat{L}u\equiv \ddot{u}+2\zeta \dot{u}+\left( \sigma _{0}+\sigma _{1}\cos
\left( \Omega t+\theta \right) +\sigma _{2}\cos \left( 2\left( \Omega
t+\theta \right) \right) \right) u=0  \label{Hill-1a} \\
\sigma _{0}=3\gamma A_{0}^{2}+\dfrac{3}{2}\gamma A_{1}^{2},\quad \sigma
_{1}=6\gamma A_{0}A_{1},\quad \sigma _{2}=\dfrac{3}{2}\gamma A_{1}^{2}
\label{Hill-1b}
\end{gather}
provided that higher powers of $u$ are neglected.

To study destabilization of the $1:1$ resonance via the period doubling
scenario one puts into Eq. (\ref{HILL-1}): 
\end{subequations}
\begin{equation}
u\left( t\right) =B\cos \left( \tfrac{1}{2}\Omega t+\varphi \right) ,
\label{uu}
\end{equation}%
obtaining, after a harmonic balance method is used, a simple necessary
condition for the onset of period-doubling (a condition for non-zero $B$): 
\begin{equation}
h \left( A_{0},A_{1},\Omega ;\gamma ,\zeta \right) \equiv \left( \sigma _{0}-%
\frac{1}{4}\Omega ^{2}\right) ^{2}+\zeta ^{2}\Omega ^{2}-\frac{1}{4}\sigma
_{1}^{2}=0,  \label{condition-1}
\end{equation}
see Eq. (8.5.5) in \cite{Kovacic2011} or Eq. (4d) in \cite{Szemplinska1986}.

Equation (\ref{condition-1}) can be simplified.  We compute $A_{1}^{2}$ from Eq.(\ref{1e}) 
for $A_{0}\neq 0$ and substitute into (\ref{condition-1}) obtaining a simplified condition for the birth of period
doubling: 
\begin{equation}
\left. 
\begin{array}{l}
k\left( A_{0},\Omega ;\gamma ,\zeta ,F_{0}\right) =10\gamma
^{2}A_{0}^{6}-\gamma \Omega ^{2}A_{0}^{4}-2\gamma F_{0}A_{0}^{3}\smallskip 
\\ 
+\left( \tfrac{1}{16}\Omega ^{4}+\zeta ^{2}\Omega ^{2}\right) A_{0}^{2}-%
\tfrac{1}{2}F_{0}\Omega ^{2}A_{0}+F_{0}^{2}=0.%
\end{array}%
\right.   \label{k}
\end{equation}

\section{Differential condition for period doubling}
\label{differential} 
We are going to show that there is a differential condition that permits a further insight 
into the nature of the birth of period doubling.

We consider a quite obvious differential condition for the birth of period doubling. 
More precisely, we investigate when equations (\ref{f}), (\ref{k}) have a common (real) root 
$\left( A_{0},\ \Omega \right) $.  Suppose that the implicit function (\ref{k}) has a singular point, 
an isolated point. Then equation (\ref{f}) guarantees that this isolated point lies on the 
$1:1$ resonance amplitude-frequency curve (\ref{f}) -- a double solution of Eqs. (\ref{f}), (\ref{k}).
This corresponds to a birth of instability of the $1:1$ resonance with the creation of period doubled 
solution (\ref{u}), i.e. to a birth of period doubling.

Accordingly, we consider the following equations: 
\begin{subequations}
\label{PD-S}
\begin{align}
f\left( A_{0},\Omega ;\gamma ,\zeta ,F_{0},F\right) & =0,  \label{res} \\
k\left( A_{0},\Omega ;\gamma ,\zeta ,F_{0}\right) & =0,  \label{pd1} \\
\frac{\partial k\left( A_{0},\Omega ;\gamma ,\zeta ,F_{0}\right) }{\partial
A_{0}}& =0,  \label{pd2} \\
\frac{\partial k\left( A_{0},\Omega ;\gamma ,\zeta ,F_{0}\right) }{\partial
\Omega }& =0,  \label{pd3}
\end{align}
\end{subequations}
where Eq.(\ref{res}), equivalent to (\ref{f}), is the steady-state condition
for $1:1$ resonance, Eq.(\ref{pd1}) is the period doubling condition (\ref{k}
) and equations (\ref{pd2}), (\ref{pd3}) mean that the implicit function (\ref{k}) has a singular point. 
We show below that this singular point is an isolated point of (\ref{k}).

Acceptable solutions of Eqs.(\ref{PD-S}), i.e $\Omega >0$, $A_{0}>0$, $%
\gamma >0$, $F>0$: 
\begin{subequations}
\label{SOL}
\begin{equation}
\begin{tabular}{|l|l|l|l|}
\hline
$\Omega _{\ast }$ & $A_{0\ast }$ & $\gamma _{\ast }$ & $F_{\ast }$ \\ 
\hline\hline
$c_{1}\,\zeta $ & $c_{2}\,\dfrac{F_{0}}{\zeta ^{2}}$ & $c_{3}\,\dfrac{\zeta
^{6}}{F_{0}^{2}}$ & $c_{4}\,F_{0}$ \\ \hline
\end{tabular}%
\   \label{sol1a}
\end{equation}%
where $\zeta >0$, $F_{0}>0$ are free parameters and 
\begin{equation}
\begin{array}{ll}
c_{1}=2\sqrt{2}\sqrt{\sqrt{2}+1}\cong 4.\,395, & c_{2}=\frac{3}{4}\left( 1-%
\frac{1}{2}\sqrt{2}\right) \cong 0.\,220, \\ 
c_{3}=\frac{32}{27}\left( 7\sqrt{2}+10\right) \cong 23.\,585, & c_{4}=\frac{3%
}{8}\sqrt{2}\sqrt{16\sqrt{2}+73}\cong 5.\,186,%
\end{array}
\label{sol1b}
\end{equation}
\end{subequations}
have been computed using Maple from Scientific WorkPlace 4.0. Finally, we
compute $A_{1\ast }^{2}$ from Eqs.(\ref{1e}), (\ref{sol1a}) and check that
inequality $A_{1\ast }^{2}>0$ is fulfilled for $F_{0}>0$\ since 
$\frac{3}{2}\gamma A_{0\ast }A_{1\ast }^{2}=F_{0}-\gamma _{\ast
}A_{0\ast }^{3}=\frac{3}{4}F_{0}>0$.

Now we demonstrate that solution (\ref{SOL}) corresponds to an isolated point. 
The determinant of the Hessian matrix, computed for the function 
$k\left( A_{0},\Omega ;\zeta ,\gamma ,F_{0}\right) $ at singular point (\ref%
{SOL}), is positive: 
\begin{equation}
\det \left( 
\begin{array}{cc}
\frac{\partial ^{2}k\left( A_{0},\Omega \right) }{\partial \Omega ^{2}} & 
\frac{\partial ^{2}k\left( A_{0},\Omega \right) }{\partial \Omega \partial
A_{0}}\smallskip  \\ 
\frac{\partial ^{2}k\left( A_{0},\Omega \right) }{\partial A_{0}\partial
\Omega } & \frac{\partial ^{2}k\left( A_{0},\Omega \right) }{\partial
A_{0}^{2}}%
\end{array}%
\right) =\left( 36\sqrt{2}+18\right) \zeta ^{2}F_{0}^{2}>0,  \label{det}
\end{equation}
and this means that this is an isolated point \cite{Fikhtengolts2014}.

For example, if we choose $\gamma =0.1$ and $\ F_{0}=0.02$ then we compute
from Eqs. (\ref{SOL}) other parameters, $\zeta $, $F$, as well as $\Omega $, $%
A_{0}$, and $A_{1}$ from Eq. (\ref{1e}).
\begin{equation}
\hspace{-0.09cm}
\begin{tabular}{|l||l||l||l||l|l|l|}
\hline
$\zeta _{\ast }$ & $\gamma $ & $F_{0}$ & $F_{\ast }$ & $\Omega _{\ast }$ & $%
A_{0\ast }$ & $A_{1\ast }$ \\ \hline\hline
$0.109\,204$ & $0.1$ & $0.02$ & $0.103\,721$ & $0.479\,923$ & $0.368\,403$ & 
$0.521\,001$ \\ \hline
\end{tabular}
\label{parameters}
\end{equation}

\begin{figure}[h!]
\center
\includegraphics[width=12cm, height=8cm]{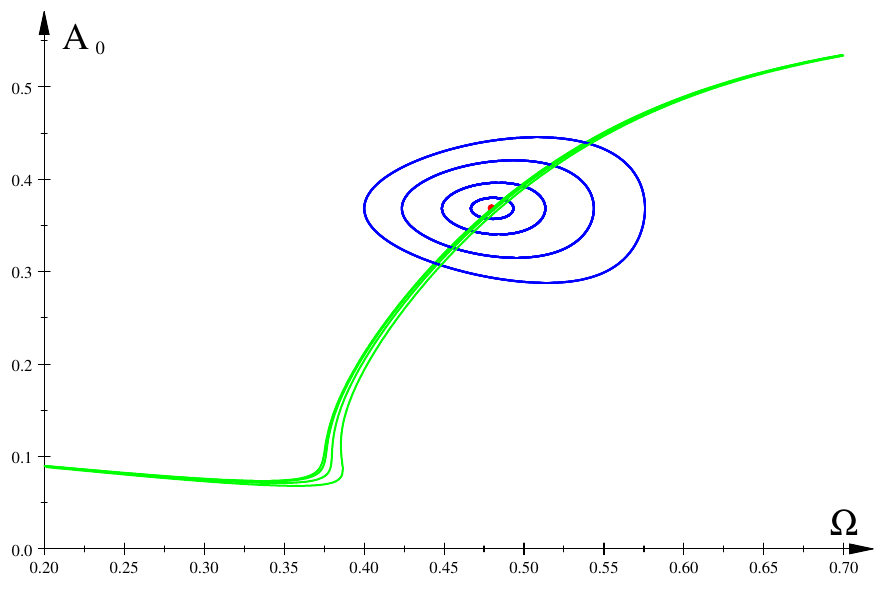}
\caption{Implicit functions: (\protect\ref{res}) (green), (\protect\ref{pd1}%
) (blue), isolated point (red); $\gamma =0.1$,\ $F_{0}=0.02$, $F=0.103\,721$ 
and $\zeta =0.109\,204$ (singular), $ 0.109$, $0.108$, $0.105$, $0.100$. }
\label{F1}
\end{figure}

And indeed, if we solve Eqs. (\ref{res}), (\ref{pd1}) for this set of
parameters we obtain a double solution $\left( A_{0\ast },\ \Omega _{\ast
}\right) =\left( 0.479\,923,\ 0.368\,403\right) $, see also Fig.\ref{F1},
where the singular point $\left( A_{0\ast },\ \Omega _{\ast }\right) $ --  an isolated point of 
implicit function (\ref{k} -- is shown as a red dot. 

For $\zeta >  \zeta_{\ast}$ solutions of Eq. (\ref{pd1}) are complex, for $\zeta = \zeta_{\ast}$ 
there is a real isolated point lying on the curve (\ref{res}) -- a red dot in Fig. \ref{F1}, and for decreasing 
values of $\zeta$ curves ($\ref{pd1}$) are growing blue ovals. 
More exactly, in Fig. \ref{F1} in all cases, we have $\gamma =0.1$,\ $F_{0}=0.02$, $F=0.103\,721$ 
while $\zeta =0.109\,204$ (singular), $ 0.109$, $0.108$, $0.105$, $0.100$.

Note that for decreasing $\zeta$ implicit function (\ref{res}), describing
the $1:1$ resonance changes only slightly while the implicit function (\ref%
{pd1}), destabilization condition of the resonance, changes significantly.

\section{Numerical verification}
\label{verification}
It follows from Section \ref{differential} that for $\gamma =0.1$,\ $F_{0}=0.02$, $F=0.103\,721$ 
destabilization of the $1:1$ resonance occurs for $\zeta \leq \zeta _{\ast }=0.109\,204$. 
Therefore, we have computed bifurcation diagrams solving Eq. (\ref{AsymDuffing})
for  $\gamma =0.1$,\ $F_{0}=0.02$, $F=0.103\,721$, and $\zeta \approx 0.109\,204$ looking for 
an onset of period doubling. And indeed, the resonance $1:1$ becomes unstable for $\zeta \cong 0.1105$. 

\begin{figure}[h!]
\center
\includegraphics[width=6.0cm, height=6.0cm]{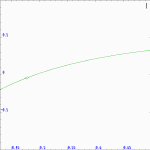} 
\includegraphics[width=6.0cm, height=6.0cm]{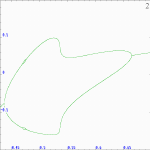} 
\vspace{0.3cm}
\includegraphics[width=6.0cm, height=6.0cm]{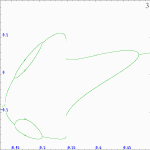}
\includegraphics[width=6.0cm, height=6.0cm]{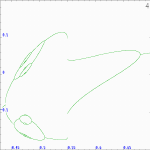} 
\caption{Bifurcation diagrams: $y$ (vertical axis) against $\Omega$ (horizontal axis), 
$\gamma =0.1$,\ $F_{0}=0.02$, $F=0.103\,721$ 
and $\zeta =0.1105$, $0.0621$, $0.0562$, $0.0550$ in diagrams $1$, $2$, $3$, $4$, respectively. 
}
\label{F2}
\end{figure}

Numerical solutions $y\left( t\right) $  of Eq. (\ref{AsymDuffing}) (bifurcation diagrams) were 
computed running DYNAMICS \cite{Nusse1997}  in the interval 
$\Omega \in \left( 0.43,\ 0.70\right) $ for $\gamma =0.1$,\ $F_{0}=0.02$, $F=0.103\,721$ and 
$\zeta =0.1105$, $0.0621$, $0.0562$, $0.0550$, see Fig. \ref{F2}.

We note that destabilization of the $1:1$ resonance 
with the formation of $1:2$ solution (\ref{PERTURBED}) (as well as other resonances) 
appears at $\Omega \approx 0.48$ in good 
agreement with the analytical value $\Omega_{\ast} = 0.479\,923$ in \ref{parameters}. 

Moreover, we have computed numerical values of parameter $\zeta$ at which the first and  subsequent 
period-doubling bifurcations occur, see Eq. (\ref{Feigenbaum}).
The first period doubling takes place at $\zeta = 0.110 \, 533 \, 4$ in good 
agreement with analytical value $\zeta_{\ast} = 0.109\,204$, see Eq. (\ref{parameters}). 
We have also computed  ratios $\dfrac{\zeta _{i-1}-\zeta _{i-2}}{\zeta _{i}-\zeta _{i-1}}$ 
which converge quite well  to the Feigenbaum constant $\delta =4.669\,2011\,609\,\ldots $  \cite{Feigenbaum1978}. 

\begin{equation}
\begin{tabular}{|c|c|c|c|c|}
\hline
$i$ & $\text{period }2^{i}$ & $\Omega _{i}$ & $\zeta _{i}$ & $\dfrac{\zeta
_{i-1}-\zeta _{i-2}}{\zeta _{i}-\zeta _{i-1}}$ \\ \hline\hline
$1$ & $2^{1}$ & $0.477\,000$ & $0.110\,533\,4$ & $-$ \\ \hline
$2$ & $2^{2}$ & $0.473\,200$ & $0.062\,204\,5$ & $-$ \\ \hline
$3$ & $2^{3}$ & $0.474\,200$ & $0.056\,225\,6$ & $8.\,083$ \\ \hline
$4$ & $2^{4}$ & $0.474\,550$ & $0.055\,066\,8$ & $5.\,160$ \\ \hline
$5$ & $2^{5}$ & $0.474\,515$ & $0.054\,824\,0$ & $4.\,772$ \\ \hline
$6$ & $2^{6}$ & $0.474\,515$ & $0.054\,771\,9$ & $4.\,660$ \\ \hline
\end{tabular}
\label{Feigenbaum}
\end{equation}

\section{Summary}
\label{summary}

Based on the known steady-state solution (\ref{AD-solution}), (\ref{1d1e}) 
and period doubling-condition (\ref{condition-1}) 
(or simplified Eq. (\ref{k})), we have computed, using the theory
of differential properties of implicit functions, a two-parameter family of singular points, see  Eqs. (\ref{SOL}), 
for implicit function (\ref{k}) -- solutions of Eqs. (\ref{PD-S}).
We have demonstrated that these singular points are isolated points which 
for $\zeta = \zeta_{\ast}$ lie on the amplitude-frequency curves of the $1:1$ resonance. 
The emergence of an isolated point corresponds to the onset of period doubling, cf.  Eqs. (\ref{parameters}), 
(\ref{Feigenbaum}).

Furthermore, we have obtained good agreement between analytical value $\zeta = \zeta_{\ast}$, Eq. (\ref{parameters}) 
and numerical value $\zeta_1$ for the onset of period doubling, Eq. (\ref{Feigenbaum}).

It is possible to control destabilization of the $1:1$ resonance by decreasing $\zeta$, $\zeta < \zeta_{\ast}$. 
Indeed, it follows from Fig. \ref{F2} and Eq. (\ref{Feigenbaum}) that upon decreasing $\zeta$ we 
observe a build-up of the Feigenbaum cascade of period doubling, leading to chaos. 
We note that all period doubling occurs for $\Omega \in \left( 0.47,\ 0.48\right) $ and this suggests 
that in the case of higher period doubling there is a similar mechanism at work.

We hope, that our approach can be applied to other periodically forced nonlinear equations.

\end{document}